\documentclass[twocolumn,showpacs,preprintnumbers,amsmath,amssymb]{revtex4}

\usepackage{graphicx}
\usepackage{dcolumn}
\usepackage{bm}
\usepackage{amssymb}

\newcommand\nuh{\nu_h \rightarrow \gamma  \nu} 
\newcommand\muh{\mu \rightarrow e  \nu_e \nu_h}
\newcommand\murad{\mu \rightarrow e  \nu_e \nu_\mu \gamma}
\newcommand\muhd{\mu \rightarrow e  \nu_e \nu_h \to e \nu_e \gamma  \nu }

\newcommand\mix{|U_{\mu h}|^2}
\newcommand\khd{K \rightarrow \mu \nu_h \to \mu \gamma  \nu }

\def\address{\@ifstar{\address@star}%
  {\@ifnextchar[{\address@optarg}{\address@noptarg}}}

\begin{document}

\author{S.N.~Gninenko}

\affiliation{Institute for Nuclear Research, Moscow 117312}


\title {The LSND/MiniBooNe excess events and  heavy neutrino from $\mu$ and $K$ decays}

\date{\today}

\begin{abstract}
It has been recently shown that puzzling excess events observed by  
the LSND and MiniBooNE neutrino experiments could be interpreted as a signal from  
the radiative decay of a heavy sterile neutrino ($\nu_h$) of the mass from 40 to 80  MeV,
with a muonic mixing strength $\mix \simeq 10^{-3} - 10^{-2}$, and the lifetime 
$10^{-11} \lesssim \tau_h \lesssim 10^{-9}$ s.
We discuss bounds on $\mix$ obtained from the recent precision measurements with muons and show 
that they are consistent with the above limits.
If the $\nu_h$ exists its admixture in the ordinary muon or kaon decay would result
in the decay chain $\muhd$ or $\khd$, respectively. We propose a new experiment for a sensitive search for 
these processes in $\mu$ and $K$  decays at rest 
allowing to either definitively confirm or exclude the existence of 
the $\nu_h$. To our knowledge, no experiment has specifically searched for 
the signature of radiative decay of massive neutrinos from particles decays
as proposed in this work. The search is complementary to the current 
experimental efforts to clarify the origin of the LSND and MiniBooNE anomalies.  
\end{abstract}
\maketitle

\section{Introduction}
Over the past 10 years there is a  puzzle of the 3.8 $\sigma$ event excess observed 
by the LSND Collaboration \cite{lsndfin}. 
This excess originally interpreted as a signal from 
$\overline{\nu}_\mu \to \overline{\nu}_e$ 
 oscillations was not confirmed by further measurements from the similar KARMEN experiment
\cite{karmen}.  
The MiniBooNE experiment, designed to examine the LSND effect, 
 also  found no evidence  for $\nu_\mu \to \nu_e$ oscillations.
However,  an anomalous  excess of low energy electronlike events
in  quasielastic neutrino events 
 over the expected standard  
 neutrino interactions  has been observed \cite{ minibnu2}.
 Recently, MiniBooNE has reported new  results from a search for
$\overline{\nu}_\mu \to \overline{\nu}_e$ oscillations \cite{minibnub}.
 An excess of events was  observed which has a small probability to be identified as  the 
background-only events. The data are found to be  consistent with $\overline{\nu}_\mu \to \overline{\nu}_e$ oscillations in the 0.1 eV$^2$
range and with the evidence for antineutrino oscillations from the LSND.

In recent work \cite{sng} (see also \cite{sng1, gg}) it has been shown  that these puzzling results could all be explained in a consistent way  by assuming  
the existence of a  heavy sterile neutrinos ($\nu_h$). The $\nu_h$ is created in  
$\nu_\mu$ neutral-current interactions and decay subsequently   into  
a photon and a lighter  neutrino $\nu$ in the  LSND and MiniBooNE detectors, 
but it cannot be produced in the KARMEN experiment  due to the high energy threshold.
The  $\nu_h$ could be Dirac or Majorana type.  The $\nu_h$ could decay  
{\em dominantly} into a $\gamma \nu$ pair  if, for example,  there is a large enough  transition
magnetic moment between the $\nu_h$ and $\nu$ mass states.  
Assuming the $\nu_h$ is produced through mixing with $\nu_\mu$, 
the combined analysis of the LSND and MiniBooNe excess events suggests that 
 the $\nu_h$  mass, mixing strength, and lifetime are,  respectively,  in the range
 \begin{eqnarray}
 40\lesssim m_h \lesssim 80~ \text{MeV},~ 10^{-3}\lesssim |U_{\mu h}|^2 \lesssim 10^{-2}, \nonumber \\
 10^{-11}\lesssim \tau_h\lesssim 10^{-9}~s.
 \label{param}
 \end{eqnarray}
A detailed discussion of consistency of these values  with the constraints from
previous searches for heavy neutrinos \cite{pdg} is presented in \cite{sng}.
 Briefly, the mixing of \eqref{param} is not constrained by the limits from the 
  the most sensitive experiments searched for extra peaks in two-body $\pi, K$ decays \cite{pdg}, because  
the $\nu_h$ mass range of \eqref{param} is 
 outside of the kinematical limits for $\pi_{\mu 2}$ decays, and  not accessible to 
 $K_{\mu 2}$ experiments due to experimental resolutions.
The parameter space of \eqref{param} cannot be ruled out by the results of high energy  neutrino experiments, 
such as  NuTeV or CHARM, as they searched for $\nu_h$'s of higher   masses ($ m_h \gtrsim 200~ \text{MeV}$)  
 decaying into muonic  final states ($\mu \pi \nu,~\mu \mu \nu,~\mu e \nu, ..$) \cite{pdg}, which are not 
 allowed in our case. The best limits on $|U_{\mu h}|^2$  derived for the mass range \eqref{param} 
  from the search for $\nu_h \to e^+ e^- \nu$ decays in the PS191 experiment \cite{ps191}, 
 as well as the LEP bounds \cite{aleph},  are found to be compatible with \eqref{param} 
 assuming  the dominance of the $\nuh$ decay. New limits on  mixing $|U_{\mu h}|^2$  obtained by 
 using the recent results on precision measurements of the muon Michel parameters by the TWIST 
experiment \cite{twist} are also found to be consistent with \eqref{param}.
 Finally, the most stringent 
bounds on $\mix$ coming from the primordial nucleosynthesis and SN1987A considerations, 
as well as the limits from the atmospheric neutrino experiments,
are also evaded due to the short $\nu_h$ lifetime.

As mentioned above, the most natural way to allow  the  radiative  decay  of heavy neutrino 
is to introduce a nonzero   transition magnetic moment ($\mu_{tr}$)  between the $\nu_h$ and $\nu$ mass states;
see e.g. \cite{moh, bovo}. Such coupling of neutrinos with photons is a generic consequence 
of the finite neutrino mass. Observations of the neutrino magnetic moment could allow us to distinguish whether neutrinos  are of the Dirac or Majorana type since the Dirac neutrinos can only have flavor conserving 
transition magnetic moments while  the Majorana neutrinos can only have a changing one.
In addition, Dirac neutrinos can have diagonal magnetic moments while  Majorana neutrinos cannot.
The nonzero magnetic moment of the neutrino, although tiny, is predicted even in the 
standard model (SM).
 The detailed calculations of the radiative neutrino decay rate in terms of 
the neutrino masses and mixings  were performed long ago,
see e.g. \cite{wil, pal, shrockdec}. 
The radiative decay mode could even be dominant, if  the $\mu_{tr}$ value is large enough; see  \cite{moh, bovo}. Originally, the idea of a large (Dirac) magnetic moment ($\gtrsim 10^{-11} \mu_B$, where
$\mu_B$ is the Bohr magneton) of the electron neutrino has been 
suggested in order to explain the solar neutrino flux variations \cite{vol}.  
Taking into account that in many extensions of the standard model the value of the $\mu_{tr}$
 is typically  proportional to the $\nu_h$ mass,  the intention  to make the radiative decay 
 of a $\nu_h \lesssim 100$ MeV 
 dominant by introducing a large transition magnetic moment (or through another mechanism)  is not particularly exotic from a theoretical viewpoint. 
Such  types of heavy neutrinos are present  in many interesting extensions of 
the standard model, such as GUT, superstring inspired models, left-right symmetric models and others, for a review; see e.g. Ref.\cite{moh}. 

The $\nuh$ decay rate due to a transition moment $\mu_{tr}$ is given by \cite{marc}
\begin{eqnarray}
\Gamma_{\nu \gamma}=\frac{\alpha}{8}\bigl(\frac{\mu_{tr}}{\mu_B}\bigr)^2 \bigl(\frac{m_h}{m_e}\bigr)^2 \bigl(1-\frac{m_{\nu}^2}{m_h^2}\bigr)^3 m_h
\label{ratemagmom}
\end{eqnarray}
where $m_\nu$($<m_h$), is the mass of the lighter  neutrino in the $\nuh$ decay, $\mu_B= \frac{e}{2m_e}$ is the Bohr magneton, 
and $m_e$ is the mass of electron. Assuming $m_\nu \ll m_h$, and $\tau_h \lesssim 10^{-9}$, which came from the requirement 
for the $\nuh$ decays to occur mostly inside the MiniBooNE fiducial volume,  results in \cite{sng}  
\begin{equation}
\mu_{tr} \gtrsim 3.7 \times 10^{-8}{\mu_B}.
\label{magmom}
\end{equation} 
A detailed discussion of the interpretation of the $\nuh$ decay in terms of transition magnetic moment and 
 consistency of the  value \eqref{magmom} with the constraints from
previous experiments is presented in \cite{sng} (see, Sec.VI.F). Let us just note, that the $\nu$ mass state should  not necessarily be  the dominant mass state  of an ordinary neutrino, 
$\nu_e,~ \nu_\mu$ or $\nu_\tau$. It could be a sterile neutrino as well. In this case,  there are no any constraints 
on the transition magnetic moment between two sterile states at all. There are also no stringent experimental bounds 
on the $\mu_{tr}$ between $\nu_h$ and $\nu$, if $\nu = \nu_\tau $, because there are  no intense $\tau$ neutrino beams. 
First experimental limits on $\mu_{tr}$ between a heavy sterile neutrino and muon neutrino 
have been obtained in Ref.\cite{gk1}, based on the idea of the 
Primakoff conversion $\nu_\mu Z \to \nu_h Z$   
of the muon neutrino into a heavy neutrino in the external Coulomb field 
of a nucleus $Z$, with the subsequent $\nuh$ decay.  
By  using the results  from the NOMAD experiment \cite{nomad2}, a 
model-independent bound $\mu_{tr}^{\mu h} \lesssim 4.2\times 10^{-8} \mu_B$ was 
set for the $\nu_h$ masses around 50 MeV (see Table 1 and Fig.2 in 
Ref.\cite{gk1}), which is also consistent with  Eq.(\ref{magmom}).  

\begin{figure}[tbh!]
\includegraphics[width=0.35\textwidth]{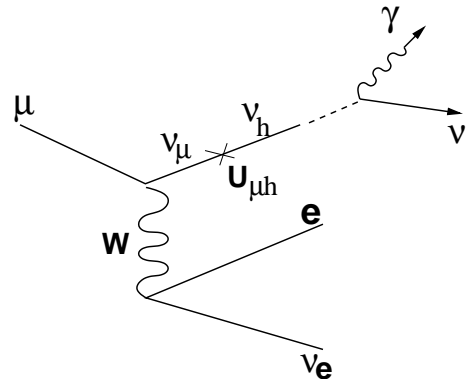}
\caption{\label{fig:diag} Schematic illustration of the production and 
subsequent radiative decay of heavy neutrinos in the ordinary muon decay. }
\label{diag}
\end{figure}  

If the $\nu_h$  is indeed  a component of $\nu_\mu$'s,  it  would be  produced by any source
of $\nu_\mu$ according to the proper mixing  and phase space and helicity factors \cite{bovo, shrock}. 
In particular, the $\nu_h$ could be produced in charge-current weak interactions of  muons or kaons.
 For example, for the mass range of \eqref{param}  the nonzero  mixing  $|U_{\mu h}|^2$
would result in the  decay $\muh$. The muon, which is normally decays 
into an $e \nu_e$ pair and a $\nu_\mu$, 
might instead decay to an $e \nu_e$ pair  and a heavy neutrino $\nu_h$ which  
decays  subsequently into $\gamma \nu$, as schematically illustrated in Fig. \ref{diag}.

In this work we show that the recent precision measurement results obtained with muons 
are consistent with \eqref{param}, and that the puzzle of the LSND-MiniBooNE excess events 
could be uniquely resolved  by the proposed  new experiment on searching  for the muon and/or kaon decay chains, $\muhd$ and $\khd$, 
respecively. 

 \section{Constraints from muon processes} 
 
For completeness of the analysis of constraints 
 reported in  \cite{sng}, let us examine first  
the recent precision  measurements results obtained with muons.  
 Note that 
for the mixing and  mass regions of \eqref{param}, the inclusion of the 
heavy neutrino effect results in the  $\muh$ decay rate which can be well approximated 
 by \cite{shrock, gs} 
\begin{eqnarray}
\Gamma(\muh)\approx \frac{G_F^2 \mix}{192\pi^3 m_\mu^3} \Bigl[m_\mu^8-m^8_h+8m^6_h m_\mu^2- \nonumber \\
-8m^2_h m_\mu^6+24 m^4_h m_\mu^4 Log(\frac{m_\mu}{m_h}) \Bigr] 
\label{rate}
\end{eqnarray} 
and  in the corresponding  branching fraction 
\begin{equation}
B(\muh) \simeq  10^{-5}-10^{-3},
\label{bratio}
\end{equation}
 which is in the experimentally accessible range. More detailed calculations 
 of the $\muh$ decay rate including arbitrary $\nu_h$ weak couplings can be found in \cite{shrockpl}.

\subsection{Muon lifetime}
Very recently, the MuLan Collaboration has reported on measurements
of the mean lifetime $\tau_\mu$ of positive muons to a precision of 0.6 ppm
\cite{mulan}. Using the new world average, 
$\tau_{\mu} = 2.1969803(22)~ \mu$s
and the relation between the muon lifetime and the Fermi constant $G_F$
$\tau_\mu^{-1}= \frac{G_F^2 m_\mu^5}{192 \pi^3}(1+\Delta)$,
where $\Delta$ is the sum of phase space, QED and hadronic corrections, 
results in  new determination of the Fermi constant 
$G_F^\mu = 1.1663788(7) \times 10^{-5}~ {\rm GeV}^{-2}$
to a precision of 0.6 ppm  \cite{mulan}.
The mixing of the $\nu_h$ into $\nu_\mu$ 
would decrease   the determined value of $G_F^\mu$.
To estimate the allowed contribution from the $\muh$ decay,  
one could compare the experimentally measured  muon decay rate 
to a predicted one, by using $G_F'$  
extracted from another measurements which are 
not directly affected  by the contribution from   heavy neutrino.  
One possible way is to use  the pure leptonic decay rate of the tau  
$\Gamma (\tau \to e \nu_e \nu_\tau)$, which provides the corresponding Fermi 
constant $G^{\tau e}_F = 1.1668 (28) \times 10^{-5}~ \rm{GeV^{-2}}$
\cite{pdg}. Comparing it with MuLan values for $G_F$ one finds 
$\Delta G_F = G^{\tau e}_F - G_F^\mu < 5 \times 10^{-3} ~GeV^{-2}~ (90\% C.L.)$.
Taking into account \eqref{rate} 
leads to the bound  $\mix <  8\times 10^{-3}$, which is consistent with \eqref{param}. One can also use a number of indirect prescriptions 
for extracting of precise values
of $G_F$ \cite{marciano}. For example, one can define 
$G_F' = \frac{4 \pi \alpha}{\sqrt{2} m_Z^2 sin^2 2\Theta_W(m_Z)(1-\Delta r)}$
where $\Theta_W$, $m_Z $  and $\Delta r$ are the Weinberg angle, the mass of 
the $Z$ gauge bosons extracted from the precision measurements at LEP, and a factor for radiative corrections, respectively.
Using the values of $\Theta_W$, $m_Z $  and $\Delta r$ reported in  
\cite{marciano}, one can obtain
$G_F' = 1.1672(\pm 0.0008)\left ( \begin{array}{c}
+0.0018 \\
-0.0007\\
\end{array} \right ) \times 10^{-5}~GeV^{-2}$.
Comparing it with $G_F^\mu$   and adding 
 statistical and systematic errors in quadrature, one finds at 90\% CL,
$\Delta G_F = G_F' - G_F^\mu < 4.1 \times 10^{-3} ~GeV^{-2}~ (90\% C.L.)$,
 which  leads to the bound  $ \mix <  7\times 10^{-3}$, which is also consistent with \eqref{param}.

\subsection{Rare muon decays}

As the final states of the decay $\muhd$ and the radiative muon decay $\murad$
are identical, the decay rate of the former  can be 
constrained  from the precise measurements of the  branching fraction  $B(\murad)$ of the later.
The most precise 
measurements of the radiative muon decay are reported by the PIBETA Collaboration \cite{pibeta}. The measured 
branching fraction $B_{exp}(\murad)= (4.4\pm 0.1)\times 10^{-3}$ is in a good agreement with the predicted value   $B_{SM}(\murad) = 4.3 \times 10^{-3}$ for the photon energy  $E_\gamma  > 15 $ MeV, and the $e-\gamma$  opening angle 
 $\Theta_{e\gamma} > 45^o$. Thus,  the  contribution from  the decay $\muh$ is allowed to be  at the level 
 $|B_{exp}(\murad)- B_{SM}(\murad)| \simeq (1.0\pm 1.0) \times 10^{-4}$, 
which, taking into account the efficiency for the above selection criteria,  
results for the $\nu_h$ mass range \eqref{param} in the 2$\sigma$ limit 
$B(\muh) \lesssim (2.6-4.1)\times 10^{-4}$, which is consistent with \eqref{bratio}. 
Interestingly, for the large values of $\Theta_{e\gamma}$, which are expected for the $\muhd$ events,  PIBETA has observed an excess of misidentified nonradiative events attributed to the  so-called  "splashback" events \cite{pibeta}. 
The size of this effect  is comparable with \eqref{bratio}.  It was assumed that these rare events 
originated probably from Michel decays  when the shower component of the decay electron 
entering an electromagnetic calorimeter
is emitted at a very large angle with respect to 
the primary electron momentum direction.

The decay chain $\muhd$ could also contribute to the background for the lepton flavor violating
 decay $\mu \to e \gamma$ whose branching fraction is constrained to be 
$B_{exp}(\mu \to e\gamma) < 1.2\times 10^{-11}$ by the MEGA experiment \cite{mega}.
In this experiment, to avoid  background from the radiative muon decay,    
only  back-to-back $e\gamma$ pairs were selected for analysis. The  
energy of the electron was required to be around the end point of   $E_e = 52.8$ MeV 
 within the energy resolution of 0.23 MeV \cite{mega}. Because the maximal allowed electron energy   
in the decay  $\muh$ is significantly less, $E_e^{max}=(m_\mu^2-m_{\nu_h}^2)/2m_\mu  < 44$ MeV,  
this potential background decay mode was  rejected by the above selection criteria. 
 
 \subsection{Muon capture}
 
Recently,  bounds on $\mix$  have been obtained from the measurements of the radiative muon capture  rate on hydrogen,
$\mu p \to \nu_\mu n \gamma$,
for the $\gamma$ energy  threshold of  $E_\gamma > 60$ MeV \cite{mcp} (see also \cite{sng2}).
The new limits seem are in tension with $\mix$ values from \eqref{param}, although 
their  exclusion strength is  a factor of few. 
 For  $\nu_h$ masses  from 40 to 70 MeV the limits are, respectively, from  
 $\mix \lesssim 7\cdot 10^{-4}$ to  $\mix \lesssim 5\cdot 10^{-4}$ (Fig. 4, $a=1$ in \cite{mcp}), 
while  the corresponding 2$\sigma$ lower bounds 
  of the allowed LSND-MiniBooNE parameter region (Fig. 24, $a=-1$ in \cite{sng}) are in the range 
   from   $\mix \gtrsim 2\cdot 10^{-3}$ to  $\mix \gtrsim 3\cdot 10^{-3}$ (In Ref.\cite{mcp} the definition of sign of
 the asymmetry parameter $a$ of the photon angular distribution in the rest frame  is  opposite to the one in \cite{sng}).
However, the  more conservative  3 $\sigma$ LSND-MiniBooNE lower bounds for the same mass range are calculated to be, 
respectively, from   
 $\mix \gtrsim 4\cdot 10^{-4}$ to  $\mix \gtrsim 5\cdot 10^{-4}$, and
the tension is not apparent. 
Furthermore , regardless of the level of the exclusion strength,
there is also a direct way to evade the radiative muon capture  limit by shifting the   photon energy 
spectrum from the $\nuh$ decay towards  the lower energy region below 60 MeV  by allowing  the neutrino $\nu$ 
to be massive enough, as suggested in \cite{mcp}.
 \begin{figure}[tbh!]
\includegraphics[width=0.45\textwidth]{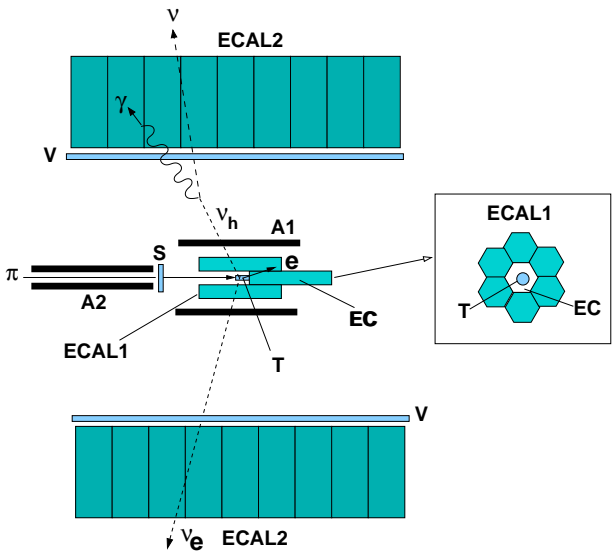}
\caption{\label{fig:setup} Schematic illustration of the 
proposed beam-dump type  experiment to search for  the decay $\mu^+ \to e^+ \nu_e \nu_h \to e \nu_e \gamma \nu $. 
Shown are the beam defining  
scintillator counter $S$,
the active target $T$ surrounded by the  
electromagnetic calorimeters ECAL1 and ECAL2, and  the 
ECAL1 endcap counter $EC$ used as a light guide for the light produced in the target.
The source of muons is the $\pi \to \mu +\nu$ decay at rest in $T$.
Electrons and photons from the ordinary muon decays $\mu^+ \to e^+ \nu_e \nu_\mu (\gamma)$ are dumped   
in the assembly, consisting of the ECAL1 and a lead absorber $A1$,  used as a shield for the ECAL2. 
Electrons and $\gamma$'s  from the $\pi,\mu$ decays in  flight before the $S$ counter are absorbed by the shield $A2$. 
The $\nu_h$  produced through the mixing  with the muon neutrino penetrates the (ECAL1+ $A1$)-assembly 
and decays in flight in the free space between the ECAL1 and ECAL2  
into a light neutrino and photon, which is detected by the ECAL2. The insert shows the front 
view of the ECAL1 area.}
\label{setup}
\end{figure}  

\section{Experiments to search for the $\nuh$ decay}

In order to (dis)prove 
the $\nu_h$ interpretation of
 the LSND-MiniBooNE excess events, we  propose a new  experiment to 
search for the $\nuh$ decay chain from the muon or kaon decays at rest with a sensitivity in $\mix$ 
 several  orders of magnitude higher than the $\mix$ values of \eqref{param}.
 
\subsection{Search for the $\muhd$ decay} 
The main components of the detector are schematically shown in Fig. \ref{setup}. A
beam of positive pions with momentum of $\simeq $ 70 MeV/c, e.g. from the $\pi$M3 pion beam line at PSI, 
, defined by a scintillator counter $S$, is stopped in an active target ($T$) instrumented with energy deposition and time readout. 
The source of muons 
is the $\pi \to \mu +\nu$ decay at rest in $T$.  
 The target $T$ is surrounded by an almost 4$\pi$ hermetic shield consisting of an  electromagnetic calorimeter ECAL1 and a lead absorber ($A1$),  
where  all the electrons and photons from the $\pi,\mu$ decays in $T$ are dumped. 
The ECAL1 is surrounded by another ECAL2 to detect 
photons from the $\nuh$  decays  in flight. It is assumed that the $\nu_h$ produced via the mixing in the $\muh$ decay
is a weakly interacting particle, which penetrates the (ECAL1+$A1$) assembly  without  significant attenuation and 
  decays subsequently in the free
space between the calorimeters, as shown in Fig. \ref{fig:setup}. Electrons and $\gamma$'s  from the $\pi,\mu$ decays in  flight before the $S$ counter, cannot hit the ECAL2, as they are absorbed by the shield $A2$. 
 The experimental signature of the decay chain $\muhd$ is two signals of the energy 
deposition in the calorimeters  separated in time  by an interval $\Delta t \lesssim 1 $ ns
corresponding to the $\nu_h$ time-of-flight. The physical background for the signal events is expected 
to be very small, because only neutrinos can penetrate the shield, 
but they  produce a negligible number of events in the ECAL2. As the proposed experiment is of the beam-dump type, there is 
no special requirements for the purity of the beam. The amount of muons and electrons from 
$\pi$ decays in flight  could be comparable to the number of pions at the end of the beam line.
  
To estimate the sensitivity of the proposed experiment 
 a feasibility study  based on simplified Monte Carlo simulations,
similar to the one described in \cite{sng3},
has been  performed. 
The pions are
stopped in the  $T$, which  is a plastic scintillator with 
a diameter of 5-10 mm and a height of 10 mm. According to simulations  the  decay muon came to rest passing about 1 mm in $T$
and depositing about 4.2 MeV in the $T$, which can be used for its identification.  The  ECAL1, is an array of 7 BGO counters, as schematically shown in Fig. \ref{setup},  each of 
55 mm in diameter and  220 mm long, which were  previously used in the PSI experiment on 
precise measurements of the $\pi_{e2}$ decay rate \cite{pienu}. The ECAL2  is an array of the same counters with electron 
energy resolution of $\simeq 2.5\%$ at 50 MeV.
The readout of the energy deposition in the  ECAL2  
is triggered by a tag signal of the electron  
appearance from the decay chain  $\pi \to \mu \to e + anything$, which is defined by a coincidence of a 
 signal from a stopped pion, a delayed signal from the  
 stopped decay muon and a delayed signal from the ECAL1.
 The light signals produced in  $T$ could be  readout through 
the  ECAL1 endcap crystal ($EC$) which acts as a light guide  as shown in Fig. \ref{fig:setup}.
The $T$ signals could be distinguished from the $EC$  signals due to their significantly 
different decay times by using the technique described 
in detail in Ref.\cite{bader}.\\
The significance of the $\nu_h$ discovery in the proposed experiment scales as
$S=2(\sqrt{n_s+n_b}-\sqrt{n_b})$, where $n_s,~n_b$ are the number of 
detected signal and background events \cite{nk}.
The number of events expected from the $\muhd$ decay chain is calculated as 
\begin{equation}
 n_{s} \simeq n_\mu B(\muh)B(\nuh) P_{\nuh} f_\gamma \epsilon_\gamma t, 
\end{equation}
where $n_\mu$ is the muon stop rate in the target, $B(\nuh)\simeq 1$ \cite{sng},
$P_{\nuh}\simeq exp(-l/c\gamma \tau_h)$ is the probability for the $\nu_h$ to decay in flight in free space
, where $l$  is an effective  (ECAL1+$A$) thickness and $\gamma \simeq 1$ is 
the gamma-factor of the $\nu_h$,
$f_\gamma \gtrsim 0.8$ is the fraction of events with the energy deposition in 
the ECAL2 $> 10$ MeV, $\epsilon_\gamma \simeq 0.1 $ is  
the average $\gamma$ detection efficiency, and $t$ is the running time.
Simulations show that the energy spectrum of decay photons is
well above $\simeq 10$ MeV for the $\nu_h$ masses of \eqref{param} and a wide 
range of the $\nu$ masses.
 To estimate $n_b$, the following  background sources are  considered:
(i)  the radiative $\pi/\mu$ decays, which have the branching fraction 
$ \simeq 10^{-2}$ for $E_\gamma \gtrsim 10$ MeV \cite{pdg}.
To suppress this background the effective thickness of the (ECAL1+$A1$)-assembly  is 
selected to be $l\gtrsim 16~X_0$  resulting in 
$\lesssim 10^{-8}$ decay photons per pion stop 
 penetrating the shield without interactions;
(ii) bremsstrahlung of decay electrons in the ECAL1 results in a leak of 
low energy photons to the ECAL2, which can be rejected by the energy 
cut  $E_\gamma \gtrsim 10$ MeV; 
(iii) accidental $\gamma$'s from  $\pi,~\mu$ radiative decays in flight are rejected by the
 shield $A2$
and  by the  requirement to have only one entering particle for a particular event. The number of $\gamma$'s from the $\murad$ decays scattered in the $S$ counter is found to be small; (iv)
accidental coincidences from cosmic rays can be removed by an active veto and 
can be neglected; (v) the neutron background is expected to be small, as the setup could be 
located at a large ($\gtrsim 30$ m) distance from the proton target.
 Finally, we found that the expected background rate $n_b \lesssim 10^{-8}/\pi$ decay is dominated 
by (i), and is well controlled  by the choice of the  absorber $A1$ thickness. 
Assuming $S\gtrsim 5$ and the muon stop rate $n_\mu \simeq  3\cdot 10^4~\mu/s$  one could expect the sensitivity in the $\muh$
 decay branching ratio as small as $B(\muh) \lesssim 10^{-9}$ for  
 the  beam exposure $\simeq$ 1 month.
 
  In the case of signal observation, the value of the $\nu_h$ mass can be evaluated from the end point of the $\gamma$-energy spectrum. To cross check  the origin of signal events,  one could use an additional shield  in  the space between 
the ECALs to absorb a fraction of photons from the $\nuh$ decays in flight, which can be well predicted. 
 If no signal events are observed, the limit on the mixing strength: 
\begin{equation}
\mix  \lesssim 10^{-8} e^{0.3/\tau_h[ns]}  \Bigl[1-\frac{m_h }{m_\mu}\Bigr]^{-7/2} 
\label{limit}
\end{equation}
could be set  as a function of the $\nu_h$ mass and lifetime. 
 Using \eqref{limit}, one can see that
 for the mass range of \eqref{param} and lifetime values in the range
 $5\cdot 10^{-11}\lesssim \tau_h \lesssim 10^{-9}$ s
 the limits on the mixing strength are $\mix \lesssim 10^{-7}  - 10^{-4}$ and hence, the $\mix$ values of  \eqref{param} 
 would be firmly excluded.
 The choice of the (ECAL1+$A1$)-assembly thickness compromises the  rejection factor of background $\gamma$'s from the $\murad$ decay 
and the sensitivity in $\mix$. For the lifetimes as short as  $\tau_h\lesssim 5\cdot 10^{-11}$ s the vast majority  of 
 $\nu_h$'s  decays in the 
vicinity of the target and the limit is less restrictive. 
To improve sensitivity for this lifetime region, one could replace 
 the ECAL1 with an assembly of thin plastic scintillator counters to detect decay electrons and the $A1$ with  an 
 absorber made of a higher-$Z$ material, e.g. tungsten, such that the  overall  thickness of the shield is reduced to a few centimeters.

\subsection{Search for the $\khd$ decay}
 
A similar  search for the $\nu_h$ can also be performed with another experiment, analogous to the one
discussed above. In this experiment we propose to search for the  $\khd$ decay chain from  kaon decays at rest.
The main component of the detector are the same as shown in Fig.\ref{setup}, but the incident beam is composed  mainly 
of positive kaons. The tag for the two-body $K\to \mu \nu$ decay is the  
energy deposition around  maximal value of 150 MeV in the ECAL1 from the stopped muon kinetic energy, which is expected for
the neutrino masses from 0 to 80 MeV. 
Similar to the above muon experiment, the experimental signature of the $\khd$ decay is two signals of the energy 
deposition in the calorimeters ECAL1 and ECAL2 separated in time  by the time interval $\Delta t \lesssim 1 $ ns
corresponding to the $\nu_h$ time-of-flight.  
The number of $K \to \mu \nu_h$ events is defined by the 
mixing $|U_{\mu h}|^2$ and by the phase space and helicity  factors which 
depend on the $\nu_h$ mass \cite{shrock1}. 
For the  mass interval $m_{\nu_h}\simeq 40-80$~MeV 
the chirality-flip is mostly due to the sterile neutrino mass  
which results in 
\begin{equation}
\Gamma(K \to \mu \nu_h) \approx 
\Gamma(K \to \mu \nu_\mu)|U_{\mu h}|^2
\rho\Bigl(\frac{m_{\nu_h}^{2}}{m_\mu^2}\Bigr),
\label{krate}
\end{equation} 
where term $\rho\Bigl(\frac{m_{\nu_h}^{2}}{m_\mu^2}\Bigr)$ includes both phase space and helicity  factors \cite{shrock1}.
Using \eqref{krate}, 
we find that the branching fraction of $K \to \mu \nu_h$ is in 
the experimentally accessible range:
\begin{equation}
Br(K\to \mu \nu_h)\simeq 10^{-3} - 10^{-2}
\label{br1}
\end{equation}  
for heavy neutrino masses in the range 40 - 80 MeV. 
Assuming the kaon stopping rate $n_K \simeq  3\cdot 10^4~\mu/s$  one could expect the sensitivity in the $K\to \mu \nu_h$
 decay branching ratio as small as $B(K\to \mu \nu_h) \lesssim 10^{-9}$ for  
 the  beam exposure $\simeq$ 1 month.
 The expected sensitivity of the kaon experiment   
 is comparable with \eqref{limit}, however 
 for the $\nu_h$ lifetime values $\lesssim 10^{-10}$ s it is even higher
 due to the higher energy of $\nu_h$'s, and, thus the longer $\nu_h$ decay length. 
This gives an advantage  for searching for the short-lived $\nu_h$
in $K$ decays  compare to the muon experiment.
  
\subsection{Search for the $\muhd$ decay with cosmic muons} 

Finally,  note that a search for  
the $\muhd$ decay events may also  be possible with existing cosmic muon data
accumulated by large Cherenkov neutrino detectors, such as SuperK \cite{sk}, MiniBooNE \cite{mbdet}, etc..
The idea is to search for  an excess of $e-\gamma$ events from stopped cosmic muon decays,
which are characterized by a clean Cherenkov 
ring from the electron from muon decay and the well separated ring from a
Compton-scattered electron from the gamma from heavy neutrino decay \cite{ref}.
The photon spectrum from the radiative muon decay drops quckly with photon energy ($\simeq 1/E_\gamma$), therefore, one would expect a low background level in the signal region of 
$E_\gamma \gtrsim 20-30$ MeV and large opening angle $\Theta_{e\gamma} \gtrsim 30^o$. However, as the  $\nu_h$ decay length ($ \lesssim 30$ cm) is significantly  less than the 
effective photon  Compton-scattering length ( $ \simeq 60$ cm), both the signal and "splashback" 
background events,   
discussed in  Sec. II B, are expected to have similar  spatial distributions with respect to the primary decay vertex.
Thus, one cannot effectively  suppress these background events by using spatial cuts. The use of cuts only on the
photon or the sum of the electron and associated photon energy deposition can not be efficient 
enough, in particular, due to a moderate energy resolution of Cherenkov detectors
in the 10 - 50 MeV energy range, e.g. $\simeq 15 \%$ at 53 MeV in MiniBooNE \cite{mbdet}. 
An estimate of the 
sensitivity of such an experiment requires  detailed simulations, which itself might  be  inaccurate in this 
extreme "splashback" regime. This is   beyond the scope of this work.
  
A search for $\muhd$ events  could also be performed  by the ICARUS \cite{icarus}. In this experiment 
the  decay electrons energy spectrum from a sample of stopping cosmic muon events recorded  
during the test run of the ICARUS T600 detector
was studied. The detector allows the spatial reconstruction of the events with fine granularity, hence, the precise measurement of the range and dE/dx of particles  with high sampling rate. 
The measured energy ($E_{exp}$) resolution for electrons below ~50 MeV is finally extracted from the simulated ($E_{sim}$) sample, 
obtaining $(E_{exp}-E_{sim})/E_{sim} = 2\%+ 11\%/E^{0.5}[MeV]$. 
 The detector has good  spatial and energy resolution, and is capable of distinguishing
 between electron and photons, hence,
 a better sensitivity  for the  search for $\muhd$ events can be expected from the analysis of recorded data.

Very recently, another interesting idea of searching for $\nu_h$'s has been proposed \cite{masip}.
It has been noticed  that a copious production of $\nu_h$'s from $K$ decays in flight could occur in air showers. Taking this into account, a sensitive search for signals from subsequent $\nu_h$ radiative decays in neutrino telescopes, like
ANTARES or the DeepCore in IceCube, could be performed.

\section{ Summary}

In summary,
we showed that the recently proposed explanation  of the puzzle 
 from the LSND, KARMEN and MiniBooNE experiments in terms of the radiative decay of a
40-80 MeV sterile neutrino,  could be uniquely probed by the 
proposed new experiment
on direct search for this decay in the muon and/or kaon decays at rest  with the sensitivity in 
a corresponding 
branching fraction as small as  a few parts in $10^{9}$. The
quoted  sensitivity could be obtained with the proposed   setup optimized for  
several of its properties, such as the  absorber $A1$ thickness, during  a month of data taking and would allow us to either confirm or rule out the existence of the $\nu_h$. 
Two different  designs could be implemented for muon experiment to cover the range of the lifetime  values 
$10^{-11}- 10^{-9}$.  The proposed searches are  complementary to the current experimental efforts
to clarify the origin of the excess events observed by the LSND and MiniBooNE experiments. This enhance motivation for 
the proposed experiments to be performed in the near future.
To our knowledge, no experiment has specifically searched for 
the signature of radiative decay of massive neutrinos from muon or kaon decays
as proposed in this work.  
The search for  events from  the $\nuh$ decay can also be performed by using the
 existing cosmic muon data, collected by neutrino detectors 
such as SuperK \cite{sk}, MiniBooNE \cite{mbdet}, ICARUS \cite{icarus}, as well as in neutrino telescopes \cite{masip}.


{\large \bf Acknowledgments}

I am grateful to D.S. Gorbunov for many 
useful discussions and help in calculations.


\begin{thebibliography}{99}
\bibitem{ref}
The idea of this search was proposed to the author by an anonymous
 referee.
\end{thebibliography}
\end{document}